# Feedback Scheduling: An Event-Driven Paradigm

Feng Xia, Guosong Tian
Faculty of Information Technology
Queensland University of Technology
Brisbane QLD 4001, Australia

f.xia@qut.edu.au

Youxian Sun
State Key Laboratory of Industrial Control
Technology,
Zhejiang University
Hangzhou 310027, China

## ABSTRACT
Embedded computing systems today increasingly feature resource constraints and workload variability, which lead to uncertainty in resource availability. This raises great challenges to software design and programming in multitasking environments. In this paper, the emerging methodology of feedback scheduling is introduced to address these challenges. As a closed-loop approach to resource management, feedback scheduling promises to enhance the flexibility and resource efficiency of various software programs through dynamically distributing available resources among concurrent tasks based on feedback information about the actual usage of the resources. With emphasis on the behavioral design of feedback schedulers, we describe a general framework of feedback scheduling in the context of real-time control applications. A simple yet illustrative feedback scheduling algorithm is given. From a programming perspective, we describe how to modify the implementation of control tasks to facilitate the application of feedback scheduling. An event-driven paradigm that combines time-triggered and event-triggered approaches is proposed for programming of the feedback scheduler. Simulation results argue that the proposed event-driven paradigm yields better performance than time-triggered paradigm in dynamic environments where the workload varies irregularly and unpredictably.

## Keywords
Feedback scheduling, programming, overhead, event-driven, resource efficiency, flexibility.

## 1. INTRODUCTION
A rapidly increasing number of microprocessors today are deeply embedded in various computing systems. Due to technical and economic reasons, embedded computing systems are typically resource constrained [2,9], for example, the available CPU time, memory, and communication bandwidth may be limited. Further, one processor using a real-time operating system or kernel often have to support the concurrent execution of multiple (real-time) tasks. While this support can be achieved through e.g. concurrent programming, the constraints on computing resources associated with the hardware platforms raise critical challenges to software design and programming since the programs have to be very resource efficient.

Many control applications are now built upon embedded systems, with every control task implemented as a single execution unit, e.g. a thread. Traditionally, real-time tasks that run in parallel are in most cases scheduled according to fixed-priority algorithms, e.g. rate-monotonic (RM), or dynamic-priority algorithms such as earliest deadline first (EDF). To meet stringent requirements of the changeful market, however, control systems have to be remarkably flexible allowing runtime reconfiguration. The end-user may be allowed to program control applications using special domain-specific programming languages. A natural result of system reconfiguration is that the workload will change accordingly. In resource-constrained environments, this variability of workload potentially causes uncertainty in resource availability. This uncertainty has unfortunately been accentuated by the increasingly popular applications of commercial off-the-shelf (COTS) components to real-time control systems. For instance, non-real-time operating systems such as Linux, Windows CE, and TinyOS have been adopted into control systems for reasons concerning cost and average performance. Although traditional scheduling policies such as RM and EDF are able to maximize temporal determinism when the resource is sufficient, they might perform very poorly in dynamic, resource-insufficient, uncertain environments, because they are inherently *open-loop* solutions [5]. In overload conditions, the required temporal behavior of the control programs cannot be guaranteed, which may causes degraded control performance or even system instability [3].

Recently, *feedback scheduling* [2,11,15] has been emerging as a promising approach to manage uncertain resources in computing systems and to enhance the flexibility and resource efficiency of the system in dynamic environments. Significant effort has been made on feedback scheduling of real-time control tasks in the last decade. Recent surveys on this research direction can be found in [8,13]. Typically, a feedback scheduler that functions as a resource manager is implemented as a *periodic* task running in parallel with control tasks. That is, the feedback schedulers are generally time triggered. In some circumstances, for example, [4,5], this triggering method has proved to be effective. However, the problem of how to determine the period of the feedback scheduler is yet to be addressed. It is intuitive that a smaller period enables better feedback scheduling performance with prompt responses to changes in workload, but also yields a larger computational overhead; a feedback scheduler with a larger period consumes less CPU resource, but responds more slowly to load variations. There are many situations in which it is very difficult to choose an appropriate period for the feedback scheduling task due to e.g. irregular variations in workload. Consequently, a time-triggered solution may leads to worse-than-possible feedback scheduling performance.

To explore the full potential of feedback scheduling, we attempt to use this technology to enhance the performance of software programs that are used in dynamic environment with uncertainty in resource availability. Our focus is on the behavioral design [14] of feedback schedulers, and this work is done in the context of real-time control applications. A general framework of feedback scheduling is given, along with a simple feedback scheduling algorithm. To provide necessary support for feedback scheduling, we describe the modifications over traditional implementation that should be made to control software programming. To improve the efficiency of feedback scheduling, we also suggest an event-driven paradigm that explores the benefits of both time-triggered and event-triggered approaches. The programming of both the control tasks and the feedback scheduling task will be improved to achieve better functional performance associated with the control software. The effectiveness and advantages of the proposed design methods will be demonstrated through simulations.

The rest of this paper is organized as follows. Section 2 describes the framework of feedback scheduling in the context of real-time control applications. A simple algorithm is given. Section 3 describes the programming method for control tasks that enables feedback scheduling. In Section 4, an event-driven paradigm for implementing feedback schedulers is presented. Simulations are conducted in Section 5 to evaluate the performance of the proposed design methods. Finally, Section 6 concludes the paper.

## 2. FEEDBACK SCHEDULING

As mentioned above, concurrent tasks are traditionally scheduled using open-loop schemes. Once configured at design time, they will not intentionally change scheduling parameters, e.g. task periods and deadlines, during run time. As a consequence, the computing resources are distributed among multiple tasks following a pre-determined pattern. In contrast, feedback scheduling introduces *feedback* into dynamic resource management, thus enabling closed-loop scheduling. The concept of feedback is the central element of *control*, a discipline that deals with the regulation of the characteristics of physical systems. Feedback is playing a critically important role in a lot of engineering systems [10]. It makes a system robust to external and internal disturbances and uncertainties. In conjunction with real-time scheduling, feedback enables the system to adapt its workload and resource allocation dynamically in a predictable manner and to achieve the desired temporal behavior [8]. In this way, the reliability of the (software) system can be enhanced, along with improved functional performance.

### 2.1 Framework

A general architecture of feedback scheduling is given in Figure 1 [11]. In the system multiple control tasks share the same real-time kernel. The control loops are typically independent, that is, the threads implementing these control tasks do not communicate with each other. Each control task executes a certain control algorithm, which is usually pre-designed using control theory and technology and responsible for controlling a physical system, i.e. a *process* or *plant* in control terms. For a control task, its period is by default equal to the *sampling period* of the corresponding control loop. In this context, the task period and the sampling period can be used interchangeably. Furthermore, it is generally assumed that the (relative) deadline of the task equals its period.

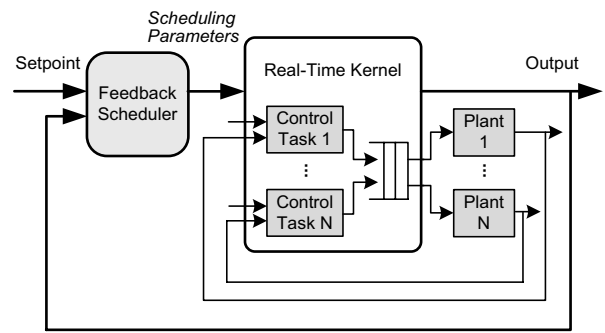

**Figure 1. A general framework of feedback scheduling**

Besides control loops, an outer feedback loop that realizing feedback scheduling is added. The basic role of the feedback scheduler is dynamically adjusting the related scheduling parameters of control tasks so as to achieve a desired resource utilization or deadline miss ratio. Feedback schedulers usually adapt the execution of a set of real-time tasks according to feedback information about actual usage of the resource. In principle, the resource referred to here can be any kind of shared resources, such as CPU time, energy, network bandwidth, among others. The feedback scheduling loop resembles a classical feedback control loop. Naturally, control theory and technology may be applied to the design of feedback schedulers. In the literature, a feedback scheduling system designed using a control theoretic approach is sometimes referred to as *feedback control (real-time) scheduling* [5]. Notice that the feedback scheduler is not actually a *scheduler* in real-time scheduling terms. It is instead a resource manager that runs on top of existing real-time scheduling policies such as RM and EDF.

To design a feedback scheduler, several variables should be chosen. In control terms, the *controlled variable* (i.e. the output of the feedback scheduling loop) and the *manipulated variable* (i.e. the scheduling parameter to be changed online) are the most important ones.

For real-time tasks, the most common performance measures are resource utilization and deadline miss ratio. The controlled variable of the feedback scheduling loop may accordingly be chosen to be either resource utilization or deadline miss ratio, as mentioned above. When the utilization is selected, the *setpoint* for the utilization should generally be maximized as much as possible to make full use of available resources, given that the system schedulability is preserved. On the contrary, when the deadline miss ratio is selected, small setpoints are preferable in order for reduction of deadline misses, given that the effectiveness of feedback scheduling is guaranteed. Both of the ranges of the possible values of resource utilization and deadline miss ratio are [0, 100%]. This indicates that in feedback scheduling systems the resource utilization and the deadline miss ratio saturate at 100% and 0, respectively, which should be taken into consideration at design time.

According to the real-time task model commonly used in real-time scheduling theory, options for the manipulated variable are the timing parameters of tasks, including period, execution time, (relative and absolute) deadlines, etc. In feedback scheduling of control tasks, the period is the most common choice for the manipulated variable. Some reasons for this are: 1) the sampling period affects simultaneously the resource demand of the control task and the resulting control performance of the loop; 2) almost all real-world control systems can perform well with a variety of sampling periods within a certain range, which makes it possible to adapt the sampling period during runtime; 3) as a design parameter in sampled-data control systems, the adaptation of sampling period is easy to realize; it is also convenient for the controller to compensate for the changes in sampling period.

Still another important issue is the overhead associated with feedback scheduling. The feedback scheduler itself consumes resources during runtime. This feedback scheduling overhead should be minimized, particularly in resource-constrained environments. The practical applicability of the feedback scheduler will be impaired if the overhead is excessively large. In time-triggered feedback scheduling, there is usually a compromise that should be made between rapid response and low overhead in order to achieve the best possible performance.

## 2.2 A Simple Algorithm

In this subsection, we will give a simple feedback scheduling algorithm. Since we do not focus on the functional design of the feedback scheduler, this algorithm will only be used in Section 5 to demonstrate the effectiveness and advantages of the proposed design methods. For this reason, we are trying to keep this algorithm as simple as possible.

Consider a multi-thread system that contains $N$ independent control tasks. Each control task $\tau_i$ is characterized by a period $h_i$ and an execution time $c_i$. For simplicity, assume both of the timing parameters are known at runtime. The number of tasks and the execution time of each task may change over time. Consequently, the requested CPU utilization, which is normally calculated by $U_{req} = \sum_{i=1}^{N} \frac{c_i}{h_i}$, will also change at runtime.

For the feedback scheduling loop, the CPU utilization and the periods are chosen as the controlled variable and the manipulated variables, respectively. Let $U_R$ denote the desired CPU utilization. Each time the feedback scheduling algorithm is activated, it will rescale all task periods using the same *period rescaling factor*, which is given by:

$$\eta = \frac{U_{req}}{U_R} \qquad (1)$$

The new period for each control task is calculated as:

$$h_i(k) = \eta \cdot h_i(k-1) = \frac{U_{req}(k)}{U_R} h_i(k-1), \quad i = 1,...,N \qquad (2)$$

If the requested utilization exceeds the setpoint, e.g. in an overload condition resulting from large increase in workload, all control tasks' utilization will be compressed through enlarging the periods. In other cases where the system is underutilized, the CPU demands of all tasks will be decompressed. After the period rescaling is performed, the total CPU utilization will return to the desired level.

It should be noted that almost all state-of-the-art feedback scheduling algorithms (e.g. [4]) applied to control systems are far more complex than this one. Consequently, these algorithms will demand significant

CPU time for each run. The feedback scheduling overhead will then become an issue that challenges the design of the feedback scheduler.

## 3. CONTROL TASK PROGRAMMING

There are different ways to implement control tasks [9,10,14]. For instance, one approach is to implement a control task as a *procedure* that is registered with the kernel to execute at a certain period and the real-time operating system will periodically call this procedure. This approach can also be supported by e.g. Real-Time Java. An alternative approach is to implement a control task as a self-scheduling task that itself contains calls to timing primitives. In practice, most periodic tasks are implemented using the second approach.

A typical implementation of a control task is shown in Figure 2 [10], where *sleepUntil* is an absolute delay primitive provided by the real-time kernel. The inner operations of the component *executeController* normally include: input sampled data via AD (Analog-to-Digital) converter, calculate control command with respect to the sampled data using pre-designed control algorithm, and output the control command to DA (Digital-to-Analog) converter.

```
Procedure Control {
   nextTime = getCurrentTime();
   while (true) {
      executeController();
      nextTime = nextTime + h;
      sleepUntil(nextTime);
   }
}
```

**Figure 2. An implementation of a control task**

When the feedback scheduler is introduced, some modifications have to be made to the programming of the control tasks. There are two main requirements to be met by these modifications. The first requirement is to support the online adjustment of task periods. A general solution for this is to augment the original inputs to the control program with a new variable, i.e., the period of the task. From a programming point of view, a shared variable representing the period will be built to facilitate communication between the feedback scheduler and each control task.

The second requirement is to compensate for the variability of sampling periods in the control algorithms. The reasons behind this requirement can be briefly explained as follows. At runtime the task periods will be adapted, implying that the sampling period of each control loop will change dynamically. From a control perspective, variations in sampling period degrade control performance, if fixed controller parameters are used. To alleviate this effect, the related controller parameters should be updated when the task/sampling period is adjusted by the feedback scheduler. For simple control algorithms such as PID (proportional-integral-derivative) and state feedback control it is possible that the controller parameters are updated directly in response to the changes in sampling periods. For complex control algorithms demanding a large amount of computations, it would be better to handle this in another way: first design offline different controllers for different sampling periods, store related controller parameters, then, during runtime, use the look-up table approach to select the most proper controller parameters for the current sampling period.

In the following, we use a PID controller, the most popular control algorithm in practical control applications, as an example to illustrate the modifications to programming induced by feedback scheduling.

```
Pre-calculation {
   bi = K*h/Ti;
   ad = Td/(M*h + Td);
   bd = M*K*ad;
}

Procedure Traditional PID Control {
   nextTime = getCurrentTime();
   while (true) {
      Input: y, r
      //y: sampled system output, r: reference input
      up = K*(r-y);
      ud = ad*ud + bd*(yold - y);
      u = up + ui + ud;
      Output: u
      //u: control command
      ui = ui + bi*(r - y);
      yold = y;
      nextTime = nextTime + h;
      sleepUntil(nextTime);
   }
}
```

**Figure 3. Traditional PID control program**

The pseudo code for traditional PID control [1,6] (without feedback scheduling) is given in Figure 3, where $K$, $T_I$, $T_D$ and $M$ are pre-set parameters. In this case, the controller parameters *bi*, *ad* and *bd* are calculated at pre-runtime and will not be updated online. Furthermore, the period is also treated as a pre-set constant.

Figure 4 shows the pseudo code of modified PID control tasks that support feedback scheduling. In contrast to traditional PID control, the period is treated as an input variable in this case. The controller parameters are updated each time the control algorithm is activated. In this way, the variations in sampling period are compensated for at the cost of slightly increased computational overhead.

```
Procedure Modified PID Control {
    nextTime = getCurrentTime();
    while (true) {
        Input: y, r, h
        //y: sampled system output, r: reference input
        //h: current period
        bi = K*h/Ti;
        ad = Td/(M*h + Td);
        bd = M*K*ad;
        up = K*(r-y);
        ud = ad*ud + bd*(yold - y);
        u = up + ui + ud;
        Output: u
        //u: control command
        ui = ui + bi*(r - y);
        yold = y;
        nextTime = nextTime + h;
        sleepUntil(nextTime);
    }
}
```

**Figure 4. Modified PID control program**

## 4. AN EVENT-DRIVEN PARADIGM

Feedback schedulers are usually time triggered, that is, they execute as periodic tasks. An obvious advantage with this mode is that it makes convenient to design and analyze the performance of the feedback schedulers using well-established feedback control theory and technology, for example, [5]. This is because sampled-data control theory in existence basically originates from periodic sampling.

The efficiency of the time-triggered mechanism can be examined in terms of response speed and overhead. To achieve quick response, feedback schedulers prefer small activation intervals so that almost all changes in workload can be treated in a timely fashion. Since the execution of feedback schedulers consumes resources, which are originally limited, the decrease of activation interval yields the increase of feedback scheduling overheads, which could adversely influence the system performance. When a relatively large activation interval is chosen for the feedback scheduler, on the other hand, it is possible that the system stays in a steady state for quite a long time, when there is actually no need for sampling period adjustment. In this situation, time-triggered feedback schedulers could potentially waste resources in periodic executions of the feedback scheduling algorithm and unnecessary updates of system parameters. Generally speaking, if the changes in workload are regular, it will be easy to determine the period for the feedback scheduling task. However, this will become very difficult when the changes in workload are irregular and unpredictable.

From the above observations, we suggest an event-driven mechanism to improve the efficiency of feedback schedulers. Discussed below is how to implement this mechanism.

### 4.1 Implementation

The schematic diagram of the event-driven activation mechanism is depicted in Figure 5. Similar to the structure of event-based controllers [1], there are basically two parts in this paradigm [12], the *event detector* and the *feedback scheduling algorithm*. The event detector is time triggered with a period of $T_{ED}$, while the feedback scheduling algorithm is triggered by the *execution-request* event issued by the event detector. This event-driven activation mechanism is generally applicable to almost all feedback scheduling methods.

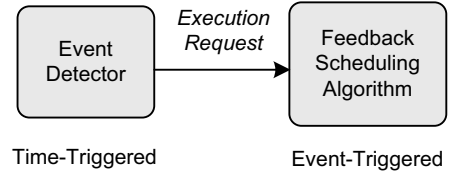

**Figure 5. Schematic diagram of event-driven paradigm**

From Figure 5, it can be seen that this paradigm is actually a combination of time-triggered and event-triggered approaches. That is why we use the term *event-driven* rather than *event-triggered*. An intuitive advantage with this event-driven paradigm is that both the predictability and flexibility inherent in time-triggered and event-triggered systems respectively can be achieved through the combination [7,14].

The key to implementing the event-driven paradigm is the design of the event detector. The major role of the event detector is deciding under what conditions the system needs to execute the feedback scheduling algorithm. As mentioned previously, the goal of the feedback scheduling loop is to maintain the CPU utilization at a desired level. Intuitively, when the utilization is in or close to steady states, there is no need for executing the feedback scheduling algorithm. On the contrary, if the utilization has significantly deviated from the desired level, then it becomes mandatory to run the feedback scheduler to adjust scheduling parameters. In this paper the following condition is used for issuing the execution-request event:

$$|U_{req}(k) - U_R| \geq \delta \qquad (3)$$

According to (3), the feedback scheduling algorithm will be executed if and only if the absolute difference between the requested utilization $U_{req}$ and its desired level $U_R$ is no less than a specific threshold $\delta$.

There are two important parameters, $T_{ED}$ and $\delta$, to be determined. Choosing these parameters generally demands careful tradeoffs between quick response and low overhead. Fortunately, this is not as difficult as usually expected. Thanks to the small amount of computations associated with (3), i.e., the even detector component, it is possible to assign a small enough period $T_{ED}$ to the event detector to achieve quick response while keeping the feedback scheduling overhead sufficiently small. The magnitude of measurement noises should be taken into account when choosing the value of $\delta$. The system usually allows the utilization to fluctuate around the setpoint with small deviations. Therefore, a $\delta$ value that is slightly bigger than the magnitude of measurement noises, for example, can be used to reduce the number of executions of the feedback scheduler, which reduces runtime overheads.

```
Procedure Time-Triggered Feedback Scheduling {
    nextTime = getCurrentTime();
    while (true) {
        executeFSAlgorithm();
        nextTime = nextTime + Tfs;
        //Tfs: period of feedback scheduler task
        sleepUntil(nextTime);
    }
}
```

(a) Time-triggered feedback scheduling

```
Procedure Event-Driven Feedback Scheduling {
    nextTime = getCurrentTime();
    while (true) {
        Calculate Ureq;
        IF (3) is true THEN
            executeFSAlgorithm();
        END IF
        nextTime = nextTime + Ted;
        sleepUntil(nextTime);
    }
}
```

(b) Event-driven feedback scheduling

**Figure 6. Different paradigms for implementing feedback scheduling task**

Figure 6 illustrates the difference between the widely-used time-triggered paradigm and our event-driven paradigm for feedback scheduling (FS). The period of the time-triggered feedback scheduler is denoted by $T_{FS}$. In a time-triggered manner, the feedback scheduling algorithm is executed every $T_{FS}$ time units. In contrast, with the event-driven paradigm, only the computations associated with the event detector are performed every $T_{ED}$ time units. The feedback scheduling algorithm will be executed if and only if necessary, i.e., when condition (3) is satisfied. Since the event detector is much less time-consuming than the feedback scheduling algorithm in most cases, it is normally possible to assign $T_{ED}$ a value much smaller than $T_{FS}$, without increasing the total CPU time consumed by the feedback scheduler. Meanwhile, the feedback scheduler will be able to respond to changes in workload more quickly. In this way, the disadvantages with time-triggered execution with respect to response speed and overhead can be avoided. Furthermore, the negative effect of measurement noises on the utilization can naturally be reduced thanks to the use of (3).

## 5. PERFORMANCE EVALUATION

In this section, we conduct simulations based on Matlab/TrueTime [6] to evaluate the performance of the above-proposed design methods for feedback scheduling. For simple description, consider three independent control loops with identical setups. The model of the controlled plant is $G(s) = 1000/(s^2+s)$, and the controller uses the PID algorithm given in Figure 4, with the following parameters: $K = 0.98$, $T_I = 0.12$, $T_D = 0.05$, and $M = 10$. The default periods of the control tasks are 10, 9, and 8 ms, respectively.

The feedback scheduling task and three control tasks are assigned fixed priorities: the feedback scheduler is given the highest priority, and the control tasks' priorities are determined by RM, i.e., $P_3 > P_2 > P_1$. The feedback scheduler uses (2) to adapt periods, and $U_R = 80\%$.

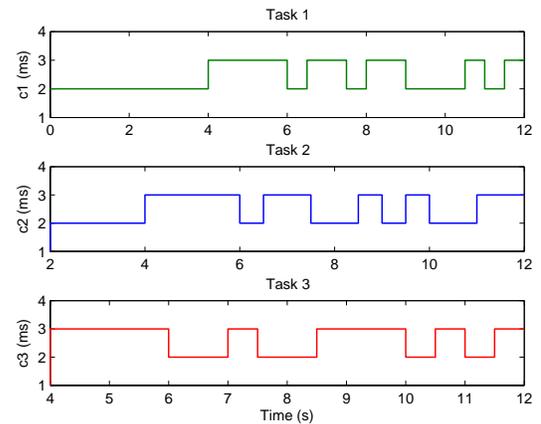

**Figure 7. Execution times of control tasks**

The simulation pattern is as follows. At time t = 0, only control task 1 is switched on. Task 2 is switched on at t = 2s. Task 3 remains off until t = 4s. The execution times of these three tasks vary at runtime according to Figure 7. This simulation pattern is simple, but sufficiently illustrative. Three different design methods are compared: 1) open-loop scheduling (OLS) that does not use any feedback scheduler, 2) time-triggered feedback scheduling (TTFS), as given in Figure 6(a), with $T_{FS} = 1s$, and 3) event-driven feedback scheduling (EDFS) given in Figure 6(b), with $T_{ED} = 0.5s$ and $\delta = 2\%$.

The integral of absolute error (IAE, a widely-used performance metric in control community) of each control loop is recorded to measure the performance of the target application. The execution time of the feedback scheduling task is set to 1ms in simulations. This relatively small value ensures that the execution of this task will not significantly impact the concurrent execution of control tasks, thus making it easy to assess the difference in control performance associated with different design methods. The overhead of feedback scheduling will be assessed roughly in terms of the *times of execution* of the feedback scheduling algorithm during a certain period of time.

## 5.1 Simulation Results and Analysis

Figure 8 depicts the sum of the IAE values of the three control loops. Notice that the larger the IAE (i.e. control cost) the worse the control performance. As expected, the open-loop scheduling method yields the worst overall performance. The system becomes unstable after time t = 4s, with rapidly increasing control cost. Since control task 1 has the low priority in the system, the first control loop will suffer most under overload conditions. As shown in Figure 9, the system becomes overloaded under OLS during time interval t = 4-6s. As a consequence, control loop 1 goes unstable, see Figure 10.

When the time-triggered feedback scheduler is used, the system remains stable during t = 4-6s, since the total CPU utilization of control tasks is kept at 80% by the feedback scheduler through period adjustment, see Figure 9. The overload occurs during this period of time under OLS is avoided, which results in satisfactory overall performance, as can be seen from Figure 10. The activation instants of the feedback scheduling algorithm are shown by the triangles in Figure 8. Since it is time triggered, the algorithm will execute every 1s. In time interval t = 0-6s, the variations in workload is regular and infrequent. Every change in workload is handled immediately. During t = 6-12s, however, the changes in workload become irregular and frequent (relative to the period of the feedback scheduling task). A considerable number of workload variations are not coped with in a timely fashion. Due to these late responses, overload conditions appear, for example, during t = 6.5-7s. The result is that the system (control loop 1) becomes unstable (Figure 10).

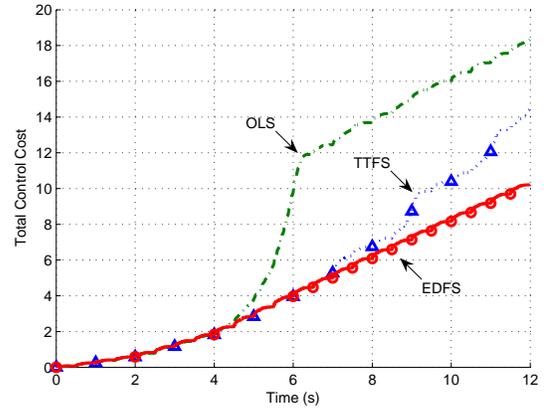

**Figure 8. Total control costs with different methods**

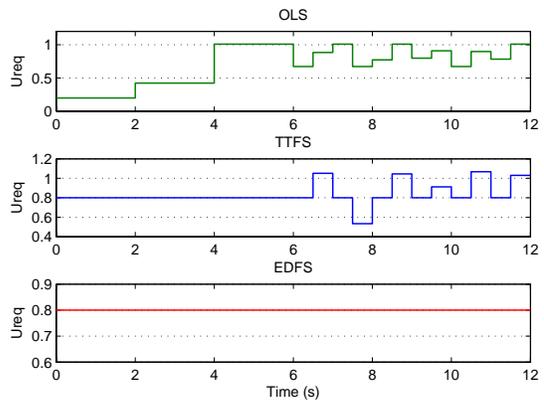

**Figure 9. Total requested CPU utilization**

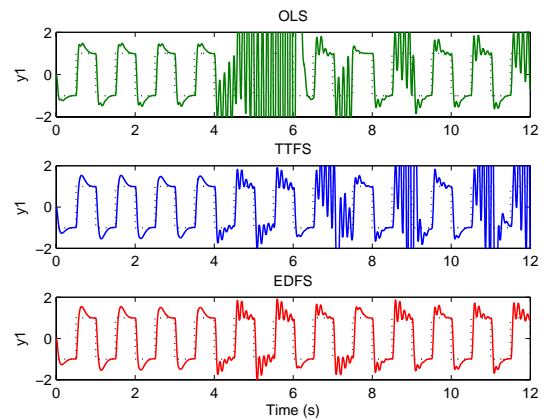

**Figure 10. System output of control loop 1**

With event-driven feedback scheduling, the system achieves the best overall performance, as shown in Figure 8, and remains stable all the time, see Figure 10. All changes in workload are handled immediately, and the CPU utilization consequently keeps at 80%, as shown in Figure 9. The activation instants of the feedback scheduling algorithm are shown by the circles in Figure 8. Compared to TTFS, EDFS yields smaller overheads in terms of times of execution of the algorithm during t = 0-6s when the workload variations are infrequent; when the workload variations become frequent, i.e. during t = 6-12s, it delivers better performance through timely responses. It can be seen that EDFS suits much better for dynamic environments in which the workload varies irregularly and unpredictably.

## 6. CONCLUSION

This paper has introduced the emerging methodology of feedback scheduling. As a promising approach to manage uncertainties in resource availability, feedback scheduling can be very useful for addressing the challenges regarding programming on resource-constrained platforms, particularly when the system operates in a variable workload environment. In the context of real-time control, we have described how to modify the programming of software/tasks to facilitate the use of feedback scheduling technology. To improve the efficiency of feedback schedulers, an event-driven paradigm that combines the benefits of both time-triggered and event-triggered approaches has also been suggested. Compared to the traditionally-used time-triggered paradigm, the proposed paradigm is more widely applicable since it can deal with not only regular but also irregular and unpredictable workload variations. When the workload varies irregularly, the event-driven paradigm performs better than the time-triggered paradigm with respect to functional performance of the program. The simulation results are also insightful for control software design and programming.

## 7. ACKNOWLEDGMENTS

The first author would like to thank Australian Research Council (ARC) for its support under the Discovery Projects Grant Scheme (grant ID: DP0559111). Part of this work was conducted when the first author was a Ph.D. candidate at Zhejiang University, China.